\documentclass[aps,prc,reprint,superscriptaddress]{revtex4-1}

\usepackage{bm}
\usepackage{longtable}
\usepackage{graphicx}

\begin{document}


\title{Fully coupled-channel study of $K^-pp$ resonance in a chiral SU(3)-based $\bar{K}N$ potential}


\author{Akinobu Dot\'e}
\email[]{dote@post.kek.jp}
\affiliation{KEK Theory Center, Institute of Particle and Nuclear Studies (IPNS), High Energy Accelerator Research Organization (KEK), 1-1 Oho, Tsukuba, Ibaraki, 305-0801, Japan}
\affiliation{J-PARC Branch, KEK Theory Center, IPNS, KEK, 203-1, Shirakata, Tokai, Ibaraki, 319-1106, Japan}

\author{Takashi Inoue}
\affiliation{Nihon University, College of Bioresource Sciences, Fujisawa 252-0880, Japan}

\author{Takayuki Myo}
\affiliation{General Education, Faculty of Engineering, Osaka Institute of Technology, Osaka 535-8585, Japan}




\date{\today}

\begin{abstract}
We have investigated the most essential kaonic  nucleus ``$K^-pp$'' as a resonant state of the $\bar{K}NN$-$\pi\Sigma N$-$\pi\Lambda N$ coupled channel system using a chiral SU(3)-based $\bar{K}N$ potential. We treat the ``$K^-pp$'' resonance adequately with a fully coupled-channel complex scaling method (full ccCSM). Self-consistency needs to be considered for the energy dependence of the chiral SU(3)-based potential. In the present study, we propose a simple prescription for the treatment of self-consistency, considering the {\it averaged threshold} and {\it averaged binding energy of mesons}. With this prescription, we have successfully found the self-consistent solutions of the ``$K^-pp$'' three-body resonance. The results indicate that the ``$K^-pp$'' system is bound rather shallowly. In particular, when the potential parameters are constrained with the latest $\bar{K}N$ scattering length, the binding energy and half of the mesonic decay width are obtained as $14-50$ MeV and $8-19$ MeV, respectively. 
\end{abstract}


\maketitle



{\it Introduction.} Finite nuclear systems with antikaons, called kaonic nuclei, have attracted interest from many researchers in the fields of strange nuclear physics and hadron physics. The interaction between antikaon ($\bar{K}$) and nucleon ($N$) is so attractive, especially in the isospin-zero channel, that it forms a quasi-bound state corresponding to the excited hyperon $\Lambda(1405)$ \cite{ChU:Review}. Previous studies based on a phenomenological $\bar{K}N$ potential, which induces a strong $\bar{K}N$ attraction, have shown a possibility that kaonic nuclei could have several exotic properties that have never been observed in ordinary nuclei \cite{AY_2002, AMDK}. Particularly, the formation of a dense state is an important property. Kaonic nuclei might be a doorway to access the dense matter, although more careful investigation should be needed. Kaonic nuclei are expected to offer hints to understand the partial restoration of chiral symmetry, which has been pursued for a long time in hadron physics \cite{ChSymRes:Hatsuda, ChSymRes:Weise}. 

Among kaonic nuclei, the most essential three-body system, $K^-pp$ (a $K^-$ meson and two protons), has been eagerly investigated from both of theoretical and experimental sides. According to theoretical studies using various approaches on the three-body system, $K^-pp$ can be bound, and its binding energy should be less than 100 MeV \cite{StrangenessSummary_2016}. However, the binding energy depends on the employed $\bar{K}N$ potential: $K^-pp$ is obtained as a shallowly bound state with a binding energy of approximately 20 MeV in the case of chiral SU(3)-based potentials (energy-dependent) \cite{Kpp:DHW, Kpp:IKS, Kpp:BGL, ccCSM+F:Dote}, whereas it is more deeply bound with a binding energy larger than 50 MeV in the case of phenomenological potentials (energy-independent) \cite{Kpp:AY, Faddeev:Shevchenko, Full-ccCSM_AY:Dote}. Furthermore, many experiments have been conducted to search for the $K^-pp$ bound state. However, their results are scattered: in some experiments, a signal around the $\pi\Sigma N$ threshold (103 MeV below the $\bar{K}NN$ threshold) was observed \cite{Kpp:exp_FINUDA, Kpp:exp_DISTO, Kpp-ex:JPARC-E27}, although there are some discussions on them \cite{Kpp_Criticism:Magas, Kpp-ex:HADES-PWA, EXA14:Gal}. In others, a signal close to the $\bar{K}NN$ threshold \cite{Kpp-ex:JPARC-E15} or no signal in the $K^-pp$ bound region was found \cite{Kpp-ex:LEPS}. Thus, a definitive conclusion on the $K^-pp$ bound state has not been achieved yet in theoretical or experimental studies. However, one experiment is noteworthy. Though the analysis is in progress, the J-PARC E15 group has reported that a signal has been clearly observed in the bound region with high statistics in the second run of their experiment of $^3$He(in-flight $K^-$, $\Lambda p$) $n_{\rm missing}$ \cite{Kpp-ex:JPARC-E15-2nd}.

In our previous study of $K^-pp$, a phenomenological potential \cite{AY_2002} was employed because such an energy-independent potential is easy to handle \cite{Full-ccCSM_AY:Dote}. In this article, we employ chiral SU(3)-based $\bar{K}N$(-$\pi Y$) potentials in order to follow the fundamental theory of strong interaction more closely. It is well known that antikaon and pion are Nambu-Goldstone bosons associated with the spontaneous breaking of the chiral symmetry of QCD at a vanishing quark mass. Pioneering works proposed a prescription to derive a $\bar{K}N$(-$\pi Y$) potential from an effective chiral SU(3) Lagrangian \cite{ChU:KSW, ChU:OR}. Subsequently, many studies have succeeded in explaining the nature of meson-baryon systems involving NG bosons, especially, the $\bar{K}N$-$\pi\Sigma$ system and $\Lambda(1405)$ \cite{ChU:Review}. 

As a method to treat $K^-pp$, we employ a fully coupled-channel complex scaling method (full ccCSM), which was developed in our previous study \cite{Full-ccCSM_AY:Dote}. All previous theoretical studies \cite{StrangenessSummary_2016} have suggested that $K^-pp$ is a resonant state that exists between the $\bar{K}NN$ and $\pi\Sigma N$ thresholds, and that it is a coupled-channel system of $\bar{K}NN$, $\pi\Sigma N$, and $\pi\Lambda N$. Full ccCSM can deal with two aspects of resonance and coupled channels adequately since it is based on the complex scaling method, which is a powerful tool for the study of resonances of many-body systems \cite{CSM:Myo}. Full ccCSM has another merit; we can reveal the structure of $K^-pp$ because this method provides us with the explicit wave function of the $K^-pp$ resonance. 

We comment on the comparison of full ccCSM with Faddeev-Alt-Grassberger-Sandhas (Faddeev-AGS) and variational methods, which were often used in past studies. As mentioned above, the coupled-channel and resonance aspects are important in the study of $K^-pp$. In Faddeev-AGS approach \cite{Kpp:IKS, Faddeev:Shevchenko}, both aspects are considered correctly, since all channels of the $K^-pp$ are treated explicitly and its resonance pole is searched for on the complex energy plane. However, only the separable-type potentials are applicable there. The wave function of the $K^-pp$ has not been shown in all past studies with this approach. On the other hand, in variational method \cite{Kpp:DHW, Kpp:BGL, Kpp:AY}, the property of the $K^-pp$ has been investigated in detail, since its wave function can easily be obtained. In principle, various types of potentials can be used in this method. However, the $K^-pp$ is calculated as a bound state of $\bar{K}NN$ with an effective $\bar{K}N$ potential in which $\pi Y$ channels are renormalized. Therefore, the coupled-channel aspect is partially considered and the resonance aspect is missed due to the bound state approximation. By the way, the present method, full ccCSM, involves both merits of two approaches, as described in the previous paragraph. However, in the full ccCSM the computational cost is obviously rather high, compared with variational approach, because of full treatment of all possible channels. Furthermore, when energy-dependent potentials are used as in the present study, self-consistency for the energy of a subsystem of the $K^-pp$ needs to be considered. To find a self-consistent solution, the calculation is repeated many times as explained in the section of methodology. In spite of high computational cost, full ccCSM is expected to be the most suitable method for the $K^-pp$ study, since it involves many advantages. 

This article is organized as follows: After the introduction, the methodology of the current study is explained. Next, the results and discussion are presented. At the end of this article, a summary of the current study and future prospects are described.


{\it Methodology.} In the present study, the $K^-pp$ system is considered as a three-body coupled-channel system of $\bar{K}NN$-$\pi\Sigma N$-$\pi\Lambda N$ with quantum numbers of spin-parity $J^\pi=0^-$, total isospin $T=1/2$ and its projection $T_z=1/2$, similarly to earlier studies. Hereafter, such a three-body state is denoted symbolically as ``$K^-pp$'' ($K^-pp$ with double quotes). In this article, ``$K^-pp$'' is investigated with full ccCSM as mentioned above. The setup for the study of ``$K^-pp$'' is the same as that of our previous study \cite{Full-ccCSM_AY:Dote}, except for the $\bar{K}N$(-$\pi Y$) potential (a potential for the $\bar{K}N$ system coupling with the $\pi Y$ system). 

The ``$K^-pp$'' wave function is represented as 
\begin{eqnarray}
|\Phi_{``K^-pp"}\rangle & = &  \sum_{ch=1}^8 \sum_{n=1}^N C^{(ch)}_n F^{(ch)}_n (\bm{x}_1, \bm{x}_2)
|S_{B_1 B_2 (ch)}=0\rangle \nonumber \\ 
&& \times \, |(M B_1 B_2)_{(ch)}; T=1/2, T_z=1/2 \rangle. \label{Kpp-wfn}
\end{eqnarray}
Here, all possible channels of $\bar{K}NN$, $\pi\Sigma N$, and $\pi\Lambda N$ are explicitly treated as indicated by labels $(M B_1 B_2)_{(ch)}$, which appear in the second line of the above equation (isospin-flavor wave function). As a result of the anti-symmetrization for baryons $B_1$ and $B_2$, the ``$K^-pp$'' wave function is composed of eight channels ($ch=1, \ldots, 8$), as listed in Table I of Ref. \cite{Full-ccCSM_AY:Dote}. The spatial wave function $F^{(ch)}_n (\bm{x}_1, \bm{x}_2)$ is represented with the correlated Gaussian basis functions \cite{CG:Suzuki} in which three types of Jacobi-coordinate sets $\{\bm{x}_1, \bm{x}_2\}$ are taken into account. Spin of baryons in each channel ($S_{B_1 B_2 (ch)}$) is assumed to be zero by following all earlier studies \cite{StrangenessSummary_2016}. The complex parameters $\{ C^{(ch)}_n \}$ are determined by the diagonalization of the complex-scaled Hamiltonian matrix as explained in a latter section.

For the ``$K^-pp$'', which is a meson-baryon-baryon system, the non-relativistic Hamiltonian $\hat{H}_{M B_1 B_2}$ is composed of the mass term $\hat{M}$, kinetic energy term $\hat{T}$, nucleon-nucleon ($NN$) potential $\hat{V}_{NN}$, and meson-baryon ($MB$) potential $\hat{V}_{MB}$: $\hat{H}_{M B_1 B_2}=\hat{M}+\hat{T}+\hat{V}_{NN}+\hat{V}_{MB}$. Here, the Argonne v18 realistic potential \cite{Av18} is employed as an $NN$ potential. As a $\bar{K}N$(-$\pi Y$) potential that is a part of $\hat{V}_{MB}$, a chiral SU(3)-based potential is employed. The details of this potential will be presented later. The $\pi N$ and $YN$ potentials are ignored in our study since their contributions are considered marginal compared to those of the $NN$ and $\bar{K}N$(-$\pi Y$) potentials. It is noted that the mass term involves the masses of all particles: $\hat{M}=\hat{m}_M + \hat{M}_{B_1} + \hat{M}_{B_2}$, where $\hat{m}_M$ and $\hat{M}_{B_i} \, (i=1,2)$ represent the meson- and baryon-mass operators, respectively. 

For the above-mentioned Hamiltonian, we calculate resonance poles of the ``$K^-pp$'' by means of complex scaling method (CSM). In this paragraph, the essence of CSM is explained briefly. More detailed explanations are given in Refs.~\cite{CSM:Myo, CSM:Myo2}. In the CSM, all coordinates \{$\bm{x}_i$\} in Hamiltonian $\hat{H}$ and wave functions $\Phi$ are transformed as $\bm{x}_i \rightarrow \bm{x}_i e^{i\theta}$ and their conjugate momenta \{$\bm{p}_i$\} are transformed as $\bm{p}_i \rightarrow \bm{p}_i e^{-i\theta}$, where the angle $\theta$ is called the scaling angle. By using the above transformation of the complex scaling, resonance wave functions, which diverge originally in asymptotic region, are modified to damping functions with appropriate values of $\theta$. Therefore, complex-scaled wave functions of resonant states ($\Phi_R^\theta$) can be described with $L^2$-integrable basis functions such as Gaussian functions, similarly to those of bound states. In addition, there is an important theorem in the CSM, which is called {\it ABC theorem} \cite{CSM:ABC-1,CSM:ABC-2}: For the complex-scaled Hamiltonian $\hat{H}^\theta$, eigen energies of resonant and bound states are independent of scaling angle $\theta$, while the energies of continuum states ($E_c$) are transformed as $E^\theta_c=E_c e^{-2i\theta}$ in CSM within non-relativistic kinematics. Namely, their energies appear on a line satisfying the equation $\tan^{-1}({\rm Im} \, E^\theta_c / {\rm Re} \, E^\theta_c)=-2\theta$ on a complex energy plane, which is called {\it $2\theta$ line}. In summary, when the complex-scaled Hamiltonian matrix is diagonalized with Gaussian basis functions, poles of resonant and bound states are stably obtained against variation of the scaling angle $\theta$, whereas energies of continuum states are obtained along $2\theta$ line. As a result, resonance poles can be identified among many complex eigenvalues of $\hat{H}^\theta$. 

Here, we make a remark on physical quantities for resonant states. The quantities such as inter-particle distances and a norm of each channel in a coupled-channel system can represent the property of the system. In CSM, we calculate the complex-scaled matrix elements as $\langle \tilde{\Phi}_R^\theta | \hat{O}^\theta |\Phi_R^\theta \rangle$ using the resonance wave function $\Phi^\theta_R$ with the complex-scaled operator $\hat{O}^\theta$, in which $\tilde \Phi^\theta_R$ represents the bi-orthogonal state of $\Phi^\theta_R$ \cite{CSM:Myo}. (The operator $\hat{O}$ corresponds to a physical quantity.) For resonant states, these matrix elements are also independent of the scaling angle $\theta$, because physical quantities should be independent of the parameter $\theta$ \cite{CSM:Homma}. In general, the quantities associated with resonances are obtained as complex values. Although the interpretation of their imaginary part is still an open problem, it is considered that the real part can represent the physical meaning as usual bound states when the magnitude of imaginary part is small \cite{CSM-meaning:Berggren, ChU:EMsize}. In this article, the CSM is applied to a coupled-channel system of the ``$K^-pp$'', and its resonance poles are obtained under the correct boundary condition.  

In our previous study \cite{ccCSM-KN_NPA}, we proposed chiral SU(3)-based potentials with a local Gaussian form factor in coordinate space. Our $\bar{K}N$(-$\pi Y$) potential is constructed from the Weinberg-Tomozawa term in the effective chiral Lagrangian, in a manner similar to coupled-channel chiral dynamics \cite{ChU:KSW} or the chiral unitary model \cite{ChU:OR}. In this article, the simplest version of our potentials is used, which is referred to as the {\it NRv2c} potential (see Eq. (8) in Ref. \cite{ccCSM-KN_NPA}).  The basic structure of this potential is 
\begin{equation}
V^{MB}_{\alpha \beta}(r, \, \sqrt{s}_{MB}) \; \propto \; -\frac{C^I_{\alpha \beta}}{8 f^2_\pi} \; (\omega_\alpha+\omega_\beta) \; g^I_{\alpha \beta}(r),  \label{Chiral_pot}\end{equation}
where $\alpha$ and $\beta$ indicate the $\bar{K}N$, $\pi \Sigma$, and $\pi \Lambda$ channels; $C^I_{\alpha \beta}$ are Clebsh-Gordan coefficients in SU(3) algebra; and $g^I_{\alpha \beta}(r)$ is a normalized Gaussian function. The meson energy in the channel $\alpha$ ($\beta$), $\omega_\alpha$ ($\omega_\beta$), is evaluated with the meson-baryon energy $\sqrt{s}_{MB}$, which is denoted as the $MB$ energy. It is noted that the meson energy term is attributed to the chiral dynamics. The pion decay constant $f_\pi$ is treated as a parameter in the present study, because our potential employs only the Weinberg-Tomozawa term in the effective chiral Lagrangian. It is varied from 90 MeV to 120 MeV so as to cover the experimental values of pion and kaon decay constants. For each $f_\pi$ value, range parameters of Gaussian functions $g^I_{\alpha \beta}(r)$ are tuned to reproduce the $\bar{K}N$ scattering length. In our previous study, the potentials, including the NRv2c potential, were determined from the $\bar{K}N$ scattering length, which was obtained by Martin's analysis of old data \cite{Exp:ADMartin}. Recently, a new value of the $\bar{K}N$ scattering length was obtained by analyzing the latest experimental result of the precise measurement of the kaonic hydrogen atom \cite{Exp:SIDDHARTA} with the chiral coupled-channel dynamics \cite{ChU:IHW}. Using the new value of the $\bar{K}N$ scattering length, we have constructed a new version of our potential, which is referred to as the {\it NRv2-S} potential. In this article, both the NRv2c and NRv2-S potentials are examined. 

As shown in Eq. (\ref{Chiral_pot}), the chiral SU(3)-based potential has an energy dependence since it explicitly includes the meson energy. To treat such an energy-dependent potential in the calculation of the ``$K^-pp$'' resonance, self-consistency for the energy needs to be considered. In this article, we follow a method to treat the self-consistency that was proposed in a previous study with a variational calculation \cite{Kpp:DHW}. The meson-baryon energy is a key quantity since it controls the energy dependence of the potential. However, the meson-baryon energy in the three-body system can not be uniquely determined in principle, because energies of partial systems are not an eigen energy of the total Hamiltonian. In the previous study of the $\bar{K}NN$ single-channel calculation \cite{Kpp:DHW}, this problem was overcome as follows. First, the antikaon's binding energy $B_K$ is calculated as a difference between the expectation values of the total $\bar{K}NN$ Hamiltonian and the $NN$ Hamiltonian. Using this $B_K$, the $MB$ energy (=$\bar{K}N$ energy) is calculated in two kinds of the extreme pictures. In one picture, called {\it field picture}, antikaon is considered as a field and this field causes 100\% of antikaon's binding energy. Since such an antikaon field interacts with an nucleon, the $MB$ energy is evaluated as $\sqrt{s}_{MB}=M_N+m_K-B_K$. Here, $M_N$ and $m_K$ are the mass of the nucleon and antikaon, respectively. In the other picture, called {\it particle picture}, antikaon is considered as a particle and it is bound by each of two nucleons equally. Since the antikaon's energy per a single $\bar{K}N$ bond should be $B_K/2$, the $MB$ energy is evaluated as $\sqrt{s}_{MB}=M_N+m_K-B_K/2$. In the particle picture, 50\% of antikaon's binding energy is used in the meson-baryon energy. In the current study, these prescriptions are generalized straightforwardly for the $\bar{K}NN$-$\pi YN$ coupled-channel calculation as follows. At first the operator of the meson's binding energy, $\hat{B}_M$, is defined as 
\begin{equation}
\hat{B}_M \; \equiv \; - \left(\hat{H}_{M B_1 B_2} - \hat{H}_{B_1 B_2} -\hat{m}_M \right),
\end{equation} 
by considering the difference between the total three-body Hamiltonian $\hat{H}_{MB_1B_2}$ and its baryon-part Hamiltonian $\hat{H}_{B_1B_2}$. We remark that the Hamiltonian $\hat{H}_{MB_1B_2}$ contains the mass term of all particles as mentioned before, whereas the baryon-part Hamiltonian $\hat{H}_{B_1 B_2}$ contains that of baryons, $\hat{M}_{B_1}+\hat{M}_{B_2}$. Therefore, the meson mass operator $\hat{m}_M$ is involved in the above equation. Using this operator and the total wave function $\Phi_{``K^-pp"}$ which is given in Eq. (\ref{Kpp-wfn}), the energy of a partial meson-baryon system in the $\bar{K}NN$-$\pi YN$ system is evaluated in two different ways as proposed in the previous study. In the first case, called the field picture ansatz, the $MB$ energy is calculated as 
\begin{equation}
\sqrt{s}_{MB} \; = \; \langle \Phi_{``K^-pp"}| \, \hat{M}_B + \hat{m}_M - \hat{B}_M \, |\Phi_{``K^-pp"} \rangle, \label{MBenergy}
\end{equation}
where the operator $\hat{M}_B$ represents a baryon mass. In the other case, called the particle picture ansatz, the term $\hat{B}_M$ in Eq.~(\ref{MBenergy}) is replaced with $\hat{B}_M/2$. Thus, the prescription used in the single-channel calculation is generalized to the coupled-channel case by considering the {\it averaged binding energy of mesons} $\langle \hat{B}_M \rangle$ together with the {\it averaged threshold} $\langle \hat{M}_B + \hat{m}_M \rangle$. In the current prescription, all quantities are averaged with the amplitudes of the coupled-channel wave function. We carry out a self-consistent calculation in terms of this $MB$ energy: Once the $MB$ energy is assumed ($\sqrt{s}_{MB, \,In}$), the $\bar{K}N$(-$\pi Y$) potential is determined. With the Hamiltonian including this potential, full ccCSM calculation is carried out to find a resonance pole of ``$K^-pp$''. According to the above-mentioned prescription, the $MB$ energy is evaluated by using the obtained resonance wave function ($\sqrt{s}_{MB, \,Out}$). Accomplishment of self-consistency is judged by comparison of these assumed and evaluated $MB$ energies. In other words, when the condition $\sqrt{s}_{MB, \,In}=\sqrt{s}_{MB, \,Out}$ is satisfied, a self-consistent solution is found to be obtained. It is remarked that the $MB$ energy is a complex value for resonant states. Self-consistent solutions are searched for in terms of the complex-valued $\sqrt{s}_{MB}$ as conducted in our previous study \cite{ccCSM+F:Dote}.


{\it Result and discussion.} The setup of the full ccCSM calculation is as follows. The spatial wave function is expanded with 20 Gaussian basis functions, whose width parameters range from 0.1 fm to 20 fm, for each relative coordinate in all sets of Jacobi coordinates. Thus, the total dimension of the complex-scaled Hamiltonian matrix, which is diagonalized to obtain resonance poles, is 6400. The scaling angle $\theta$ in the present ccCSM calculation is set to 22$^\circ$. At this angle, it is found that physical quantities like composition of the ``$K^-pp$'' and mean distance between particles as well as its resonance energy are most stably obtained against the $\theta$ variation. In this article, all energies and lengths are given in units of MeV and fm, respectively.

\begin{figure}[t]
\includegraphics[width=0.43\textwidth]{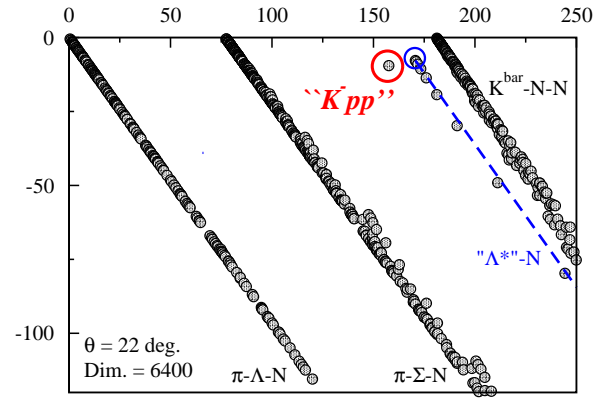}%
\caption{(Color online) Distribution of the complex-energy eigenvalues when the self-consistent solution for $``K^-pp"$ is obtained. The NRv2c potential ($f_\pi=110$ MeV) is used and the field picture is employed as the ansatz for self-consistency. The horizontal axis shows the real energy measured from the $\pi\Lambda N$ threshold, and the vertical axis shows the imaginary energy. The eigenvalues marked with red and blue circles correspond to the poles of the three-body $``K^-pp"$ resonance and the two-body $\bar{K}N$-$\pi\Sigma$ resonance, respectively. The thresholds of $\pi\Lambda N$, $\pi\Sigma N$, and $\bar{K}NN$ are located at 0 MeV, 77 MeV, and 181 MeV on the horizontal axis, respectively. \label{Fig_NRv2c-fp110-F}}
\end{figure}

As shown in Fig.~\ref{Fig_NRv2c-fp110-F}, the eigenvalues of the complex-scaled Hamiltonian are distributed on the complex energy plane where the self-consistent solution of ``$K^-pp$'' is obtained. The distribution of the complex eigenvalues is similar to that obtained in our previous study with a phenomenological potential \cite{Full-ccCSM_AY:Dote}. Eigenvalues along three lines starting from $\pi\Lambda N$, $\pi\Sigma N$, and $\bar{K}NN$ thresholds represent three-body continuum states of $\pi\Lambda N$, $\pi\Sigma N$, and $\bar{K}NN$, respectively (continuum states are known to appear along the $2\theta$ line as explained in the previous section). In addition, the two-body continuum states of $\Lambda^* N$ appear along another line. The eigenvalue other than those along these $2\theta$ lines corresponds to the pole of the ``$K^-pp$'' three-body resonance. The binding energy of ``$K^-pp$'' ($B_{K^-pp}$) and half of the mesonic decay width ($\Gamma_{\pi YN}/2$) are found to be 23.4 MeV and 9.5 MeV, respectively, when the NRv2c potential ($f_\pi=110$ MeV) and the field-picture ansatz are employed. When the other ansatz, the particle picture, is employed, $B_{K^-pp}$ and $\Gamma_{\pi YN}/2$ are obtained as 36.4 MeV and 13.4 MeV, respectively. 

\begin{figure}[t]
\includegraphics[width=0.4\textwidth]{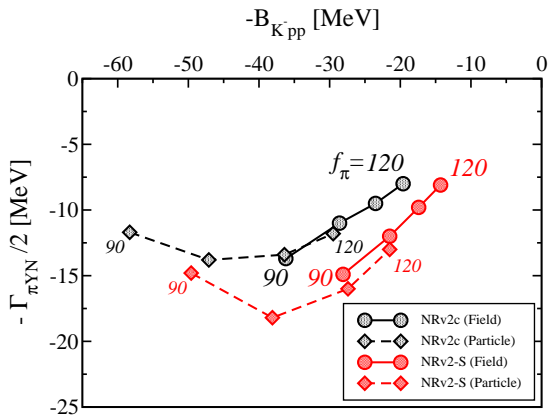}%
\caption{(Color online) Pole energy of the self-consistent solution of the $``K^-pp"$ resonance (measured from the $\bar{K}NN$ threshold). Results obtained with the NRv2c potential and NRv2-S potential are depicted in black and red, respectively. A parameter $f_\pi$ of the potentials is varied from 90 MeV to 120 MeV with 10-MeV intervals. For each $f_\pi$ values, the range parameters of Gaussian form factors in the potential (Eq.~(\ref{Chiral_pot})) are adjusted to reproduce the experimental value of the $\bar{K}N$ scattering length. The solid line with circles and dashed line with diamonds indicate the field and particle pictures, respectively. \label{Fig_Kpp_results}}
\end{figure}

In the present calculations, both the NRv2c and NRv2-S potentials with various $f_\pi$ values are used under the two ansatzes of field and particle pictures. As summarized in Fig.~\ref{Fig_Kpp_results}, the binding energy and half of the mesonic decay width are obtained as 
\begin{equation}
(B_{K^-pp}, \Gamma_{\pi YN}/2) \; = \; (19-59,\, 8-14) \; {\rm MeV} 
\end{equation}
with the NRv2c potential and 
\begin{equation}
(B_{K^-pp}, \Gamma_{\pi YN}/2) \; = \; (14-50,\, 8-19) \; {\rm MeV} 
\end{equation}
with the NRv2-S potential. In both ansatzes, a self-consistent solution is successfully obtained within the range of $f_\pi$ values. As shown in the figure, a larger binding energy and broader width are obtained for the potentials with a smaller $f_\pi$ value. In the results obtained with the field picture, it is found that a linear correlation exists between the binding energy and decay width. Compared with the field picture, the particle picture gives a larger binding energy. 

By comparing these results, ``$K^-pp$'' is more shallowly bound with the NRv2-S potential than with the NRv2c potential. The binding energy is found to be at most 60 MeV in the case of the NRv2c potential and 50 MeV in the case of the NRv2-S potential. It is emphasized that the latter potential is constrained by the latest value of the $\bar{K}N$ scattering length, as stated in the previous section. The decay width is found to be almost the same between these two potentials. Similarly to previous studies \cite{Kpp:DHW, ccCSM+F:Dote}, the decay width in the field picture is approximately half of that in the particle picture, if they are compared at the same $f_\pi$ value. 

We comment on the ansatz for the treatment of energy-dependent potentials. According to our previous study, in which the energy dependence attributed to the elimination of the channels except for $\bar{K}NN$ was investigated, the field picture was found to provide a more reasonable solution than the particle picture \cite{Full-ccCSM_AY:Dote}. If we focus on the results obtained with the field picture, the binding energy is restricted to be rather small: ($B_{K^-pp}$, $\Gamma_{\pi YN}/2$) = ($19-37$, $8-14$) MeV and ($14-28$, $8-15$) MeV with the NRv2c and NRv2-S potentials, respectively.

The properties of ``$K^-pp$'' can be investigated using the full ccCSM wave function. As a typical example, we show the $f_\pi=110$ MeV case for both potentials. In the case of the NRv2c potential, the mean distance between two nucleons is obtained as $2.10-0.06i$ fm ($1.81+0.09i$ fm) with the field (particle) picture. The NRv2-S potential gives a similar value of $2.14-0.16i$ fm ($1.80-0.06i$ fm) with the field (particle) picture. In both potentials, since ``$K^-pp$'' is obtained as a more shallowly bound state with the field picture than with the particle picture, the $NN$ distance with the former ansatz is larger than that with the latter ansatz. It is noted that the mean $NN$ distance is 2.2 fm in nuclear matter with normal density ($\rho_0=0.16$ fm$^{-3}$). Therefore, when the field picture is employed, the mean $NN$ distance of ``$K^-pp$'' is almost the same as that of the normal nuclear matter, as reported in previous studies using chiral SU(3)-based $\bar{K}N$ potentials \cite{Kpp:DHW, Kpp:BGL, ccCSM+F:Dote}. 
Table \ref{Prop_K-pp_NRv2c-110-F} lists the results obtained with the NRv2c potential with the field picture as a typical case. As indicated in the table, the $\bar{K}N$ distance in the isospin $I=0$ channel is smaller than that in the isospin $I=1$ channel, and it is close to that of the $I=0$ $\bar{K}N$-$\pi\Sigma$ resonance ($\Lambda^*$). The norm of each component indicates that, in ``$K^-pp$'', the $\bar{K}NN$ component is dominantly contained, the $\pi\Sigma N$ component is significantly mixed, and the $\pi\Lambda N$ component is very minor. Such a composition is similar to the case of using a phenomenological potential \cite{Full-ccCSM_AY:Dote}.

\begin{table}
\caption{Properties of the ``$K^-pp$'' resonant state. The NRv2c potential ($f_\pi=110$ MeV) and the field picture are used. The left column shows the mean distance between particles in the $\bar{K}NN$ channel. ``$R_{\bar{K}N(I=0,1)}$'' is the isospin-decomposed $\bar{K}N$ distance. ``$R_{\bar{K}N(I=0); \Lambda^*}$'' is the $\bar{K}N$ distance of the $\Lambda^*$ resonance. The right column shows the norm of each component in ``$K^-pp$'', ``${\cal N}(\bar{K}NN)$'' denotes the norm of the $\bar{K}NN$ component, and so on.  \label{Prop_K-pp_NRv2c-110-F}}
\begin{ruledtabular}
\begin{tabular}{lrrlr}
$R_{\bar{K}N(I=0)}$ & $1.60-i0.13$  & &  ${\cal N}(\bar{K}NN)$ & $1.128-0.071i$\\
$R_{\bar{K}N(I=1)}$ & $2.22-i0.09$ & & ${\cal N}(\pi\Sigma N)$ & $-0.127+0.067i$ \\
$R_{\bar{K}N(I=0); \Lambda^*}$ & $1.37-i0.37$ & & ${\cal N}(\pi\Lambda N)$ & $-0.002+0.004i$ 
\end{tabular}
\end{ruledtabular}
\end{table}

As a reference, we show the results switching off the energy dependence of the chiral SU(3) potential. When the meson-baryon energy in the potential is artificially fixed to be $\bar{K}N$-threshold energy ($\sqrt{s}_{MB}=M_N+m_K$), the ``$K^-pp$'' resonance pole is obtained as ($B_{K^-pp}$, $\Gamma_{\pi YN}/2$) = (45.9, 53.4) MeV for NRv2c potential and (22.2, 33.5) MeV for NRv2-S potential with $f_\pi=110$ MeV. Compared with the self-consistent results, binding energy and decay width are rather large in case of NRv2c potential. However, in case of NRv2-S potential, which is constrained with the latest $\bar{K}N$ data, the binding energy remains as small as 20 MeV.


{\it Summary and future prospects.} We have investigated the most essential kaonic nucleus, ``$K^-pp$'', with a fully coupled-channel complex scaling method \cite{Full-ccCSM_AY:Dote} that correctly treats the resonance and coupled-channel aspects, which are important factors for this system. To investigate ``$K^-pp$'' along with QCD as possible, we employed a chiral SU(3)-based $\bar{K}N$(-$\pi Y$) potential. For the treatment of the energy dependence of the chiral SU(3)-based potential, we proposed a simple prescription for the coupled-channel calculation, extending the earlier prescription for single-channel calculations \cite{Kpp:DHW, Kpp:BGL, ccCSM+F:Dote}. We introduce the averaged threshold and the averaged binding energy of mesons to estimate the meson-baryon energy in ``$K^-pp$''. Using these averaged quantities, we performed the self-consistent calculation in the same manner as in previous studies. With the extended prescription, we have succeeded in obtaining self-consistent solutions of ``$K^-pp$'' resonance with the full ccCSM calculation.  

As a result, it is found that the binding energy of ``$K^-pp$'' does not exceed 60 MeV when our chiral SU(3)-based potential is constrained with the old value of the $\bar{K}N$ scattering length. When the potential is constrained with the new value of the $\bar{K}N$ scattering length, which is obtained from the latest experiment of the kaonic hydrogen atom \cite{Exp:SIDDHARTA}, the binding energy is 50 MeV at most. If we choose the field picture, which seems to be the better ansatz when considering the self-consistency for the meson-baryon energy, ``$K^-pp$'' is found to be rather shallowly bound: the binding energy is $14-28$ MeV and half of the mesonic decay width is $8-15$ MeV. The structure of ``$K^-pp$'' can be analyzed by using the ccCSM wave function. In such a shallowly bound ``$K^-pp$'', the mean distance between two nucleons is found to be 2.2 fm, which is equal to that of normal nuclear matter. For the coupled-channel effect, the $\bar{K}NN$ component is dominant, the $\pi\Sigma N$ component is significantly mixed, and the $\pi\Lambda N$ component is quite minor. These properties are similar to the case in which a phenomenological potential is used \cite{Full-ccCSM_AY:Dote}.

By combining our previous and present studies with the full ccCSM calculation, our conclusion on ``$K^-pp$'' is as follows. If the $\bar{K}N$ potential is as strongly attractive as an energy-independent phenomenological potential \cite{AY_2002}, ``$K^-pp$''  is rather deeply bound with a binding energy of about 50 MeV. Owing to the strong $\bar{K}N$ attraction, two nucleons approach each other against the repulsive core of the $NN$ potential. The mean $NN$ distance is shorter than that of normal nuclear matter \cite{Full-ccCSM_AY:Dote}. On the other hand, if the $\bar{K}N$ potential is as weakly attractive as an energy-dependent chiral SU(3)-based potential, ``$K^-pp$'' is shallowly bound with a binding energy less than 50 MeV. Two nucleons keep the same distance as in normal nuclear matter since the $\bar{K}N$ attraction is not so strong. 

As mentioned in the introduction, the J-PARC E15 group has found a signal indicating the $K^-pp$ bound state in their second run, and they are analyzing the acquired data in detail. Their result obtained from high-statistics data, together with our theoretical result, is expected to provide a clear understanding of $K^-pp$. In our study, there still remain future tasks to be considered: 1) In the current study, only the Weinberg-Tomozawa term in the effective chiral Lagrangian was employed, which is an $s$-wave interaction. We may need to include NLO interactions and so on. 2) Concerning the decay width, the two-nucleon absorption of the antikaon ($\bar{K}NN \rightarrow YN$) is known to have a sizable contribution \cite{2Nabs:Bayar-Oset}, which is not included in the present study. It will be interesting to estimate this contribution in future work. 3) For the comparison of theoretical results with the experimental results, it is better to use the reaction spectrum than the pole position when the resonance involves a large decay width. In fact, such a study has been conducted with some approximations \cite{JPARC-E15:Sekihara-Oset-Ramos}. Within the complex scaling method, the reaction spectrum can be calculated with the complex-scaled Green's function utilizing the complex-scaled eigenstates \cite{CSM:Myo2}. Using this technique, we will study the reaction spectrum of the $K^-pp$ production. Considering these future tasks, we believe that a definitive conclusion will be reached for the longstanding issue of the essential kaonic nucleus, $K^-pp$. 

{\it Acknowledgments.} One of the authors (A. D.) thanks Prof. T. Harada and Prof. Y. Akaishi for fruitful discussions and Prof. H. Horiuchi and Prof. H. Toki for their strong encouragement. This work is supported partially by Grant Numbers 15K05091, 18K03660 and 26400281. The calculation in this study was performed using the High Performance Computing system (miho) at Research Center for Nuclear Physics (RCNP) in Osaka University.  



\end{document}